# A new experimental method to study the influence of welding residual stresses on fatigue crack propagation

Deschênes, P.-A.[a], Lanteigne, J.[b], Verreman, Y.[a], Paquet, D.[b], Lévesque, J.-B.[b], Brochu, M.[a]

[a]Department of Mechanical Engineering, École Polytechnique de Montréal, Montreal (Québec), H3T 1J4, Canada

[b]Hydro-Québec, Institut de recherche d'Hydro-Québec, Varennes (Québec), J3X 1S1, Canada

***Corresponding author*** : Pierre-Antony Deschênes (pierre-antony.deschenes@polymtl.ca)

## Abstract

This paper presents a study on the influence of welding residual stresses (RS) on fatigue crack propagation rate (FCPR). The objective of this work is to develop a novel methodology that allows a variation of a RS field in the studied specimen while keeping constant all other variables influencing FCPR. This led to the development of a novel specimen geometry, named CT-RES, in which RS are introduced by weld bead deposition far from the region in which fatigue crack propagation (FCP) occurs. As a consequence, the effect of factors influencing FCPR other than RS such as microstructural changes or plastic deformation, often introduced by welding processes, can be avoided. The welding RS introduced in the CT-RES specimen were determined by the contour method and the weight functions method was used to calculate the stress intensity factor (SIF), $K_{res}$, resulting from the RS as the fatigue crack propagates into the specimen. The evolution of cyclic stress ratio at the crack tip, $R_{local}$, was then computed from $K_{res}$ to quantify the influence of RS on the cyclic stress ratio. The results show that for a given stress intensity range, $\Delta K$, the FCPR of the welded geometry with fixed externally low $R$ ratio ($R = 0.1$), but constantly increasing $R_{local}$, is the same as for the as-machined geometry without RS, solicited at high cyclic stress ratio ($R = 0.7$). These observations partially validate the BS7910 standard philosophy in which the remaining life of a flawed structure in presence of tensile RS is calculated from a high cyclic stress ratio ($R \geq 0.5$) crack propagation curve to eliminate crack closure effects.

## 1. Introduction

Francis runners used in the hydropower industry are composed of a crown (1), blades extending from the crown to the band (2), and a band (3), as shown in Figure **1**a. These steel components are assembled by welding. It was demonstrated that performing a stress relief heat treatment does not completely eliminate welding RS introduced in turbine runners during their fabrication(Lanteigne, Baillargeon, & Lalonde, 1998). Moreover, during blades repair, E309L austenitic stainless steel is often chosen as the weld material, since using martensitic stainless steel as filler material requires a post-weld heat treatment. This type of heat treatment is complex and expensive to do in a turbine environment, as during blade repairs, hence the preference for E309L. However, the blade is left with a large amount of tensile residual stresses. Thus, like the tensile residual stresses generated during the fabrication, those coming from blade repairs accentuate the need for this study.

The presence of RS in structures can have either beneficial or detrimental effects on FCPR. A number of studies have shown that compressive RS increase the fatigue life of cracked components (Beghini & Bertini, 1990; Fleck, Smith, & Smith, 1983; Ghidini, 2007; Itoh, Suruga, & Kashiwaya, 1989; Jones, 2008)while other studies have shown that tensile RS decrease it (Liljedahl, Zanellato, Fitzpatrick, Lin, & Edwards, 2010; Ohta, Maeda, Kosuge, Machida, & Yoshinari, 1989; Ohta, McEvily, & Suzuki, 1993; Trudel, Sabourin, Lévesque, & Brochu, 2014). Thus, there is a direct correlation between the orientation of the stress component normal to the propagation plane and its influence on FCPR. In those studies, RS were introduced either by a single overload in a notched geometry(Fleck et al., 1983), by plastic deformation of the sample(Jones, 2008), by localized heat treatment (Ohta et al., 1993)or by welding (Beghini & Bertini, 1990; Ghidini, 2007; Itoh et al., 1989; Liljedahl et al., 2010; Ohta et al., 1989; Trudel, Sabourin, et al., 2014). In all these studies, the effect of RS on FCPR was rationalized by its influence on fatigue crack closure, which is known to significantly affect fatigue crack propagation rates.

However, none of the above mentioned methods adequately represents the RS distribution in Francis turbine blades or their redistribution during propagation. Indeed, the welded blades are highly constrained by a clamping effect resulting from the rigidity of the crown and band. Furthermore, all studies investigating crack propagation in welding RS fields were performed in regions where microstructural modifications and/or plastic deformations were introduced by the welding process. In order to avoid those microstructural modifications and reproduce a more representative environment, an alternate method had to be developed.

The purpose of this research work is to design a fatigue crack propagation experiment in which: (i) the specimen used would replicate the clamping effect observed in actual Francis turbine blades; and (ii) the FCP would occur in a region of the specimen exempt from any microstructural transformation or plastic deformation. To the best of the authors' knowledge, such an experiment has not yet been published.

## 1.1 The CT-RES: A novel specimen geometry

The particular specimen geometry used in this study to characterize the influence of RS on FCPR was developed following three constraints that were required to reproduce realistic conditions in which cracks propagate in Francis turbine blades:

1. RS must be introduced in the fatigue test specimen by welding;
2. FCP must not be influenced by any microstructural changes induced by this welding process;
3. Weld beads must be deposited so as to avoid any plastic deformation in the region of propagation;

The second restriction was added to avoid the inevitable microstructural effects (Trudel, Lévesque, & Brochu, 2014)when a crack propagates from the vicinity of a weld toe towards the weld deposit.

A new specimen geometry was developed in order to design a fatigue crack propagation test methodology that respects all three listed constraints. A schematic of this sample, named the *Compact Tensile Residual Stress* (CT-RES) specimen, is shown in Figure **1**b. A Francis turbine is also included in the figure to emphasize on the similarity between the two geometries: the medallion acts as a turbine blade while the rigid frame stands for the clamping effect resulting from the rigidity of the crown and the band. This particular specimen geometry is obtained by welding the medallion into a rigid frame, Figure **1**b.

Figure 2 presents different views of the CT-RES specimen. The different variables defining its geometry are also shown in Table **1**. The crack length is measured with two quantities, namely $a$ and $a*$ (*cf.* Fig.2a). The crack length is quantified using as reference the loading line ($a$) and the boundary of the medallion ($a*$) respectively.

A specimen having the same geometry as the CT-RES was machined from the same heat treated block of steel, therefore without RS. This sample, named CT-Monoblock, was used as a reference with which the fatigue crack propagation rate was compared to the welded CT-RES specimen to highlight the influence of RS on FCPR.

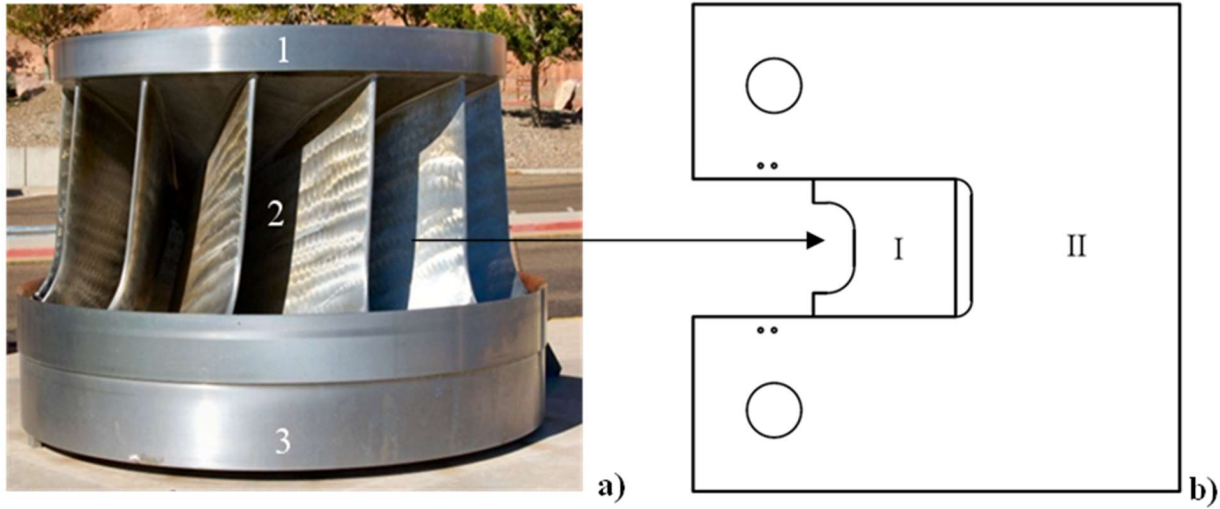

Figure 1 :(a) A typical Francis turbine composed of a crown (1), blades extending from the crown to the band (2), and a band (3) (Hall, 2005), and (b) Schematic representation of the CT-RES specimen designed to reproduce the fatigue crack behavior of Francis turbine blades. The assembly is made of a medallion (I) and a rigid frame (II)

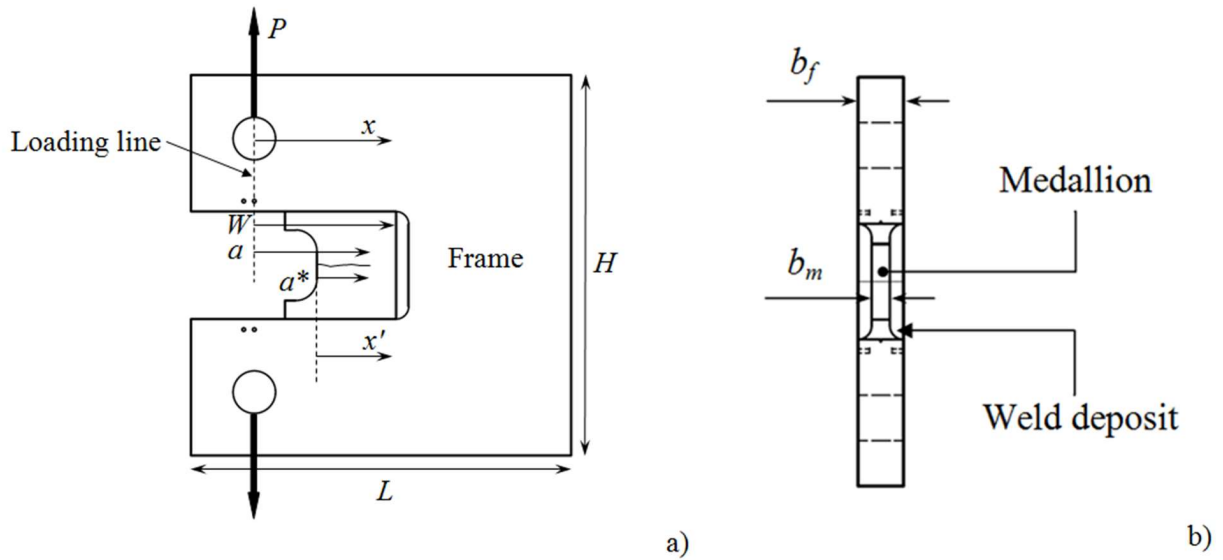

Figure 2: a) Front view of the CT-RES specimen; and b) Side view of the CT-RES specimen showing relative thickness dimensions

| $L$[mm] | $H$ [mm] | $W$ [mm] | $b_f$[mm] | $b_m$[mm] |
|---|---|---|---|---|
| 228.6 | 228.6 | 85.16 | 25.4 | 10.16 |

Table 1: CT-RES specimen dimensions

## 2. Materials properties and experimental set-up

### 2.1 Materials properties

The base metal used for the fabrication of the rigid frame and the medallion of the CT-RES specimen is UNS S41500 martensitic stainless steel. The samples were cut from Francis turbine blade segments sent by the manufacturer. Prior to machining, the material was heat treated at a temperature of 1050 C° during one hour. The material was then cooled to room temperature by slightly opening the oven door. Following this austenitization and air quenching treatment, Figure **3**, the material was further tempered at a temperature of 620 C° for two hours in order to stabilize the reformed austenite and obtain a tougher martensite. Following the heat treatment, the hardness of the steel is measured at 25 HRC. The average primary austenitic grain size was measured at 102μm by the equivalent spherical diameter method. The average percentage of reformed austenite measured by x-ray diffraction (XRD) on three different specimens was 14.8% in volume.

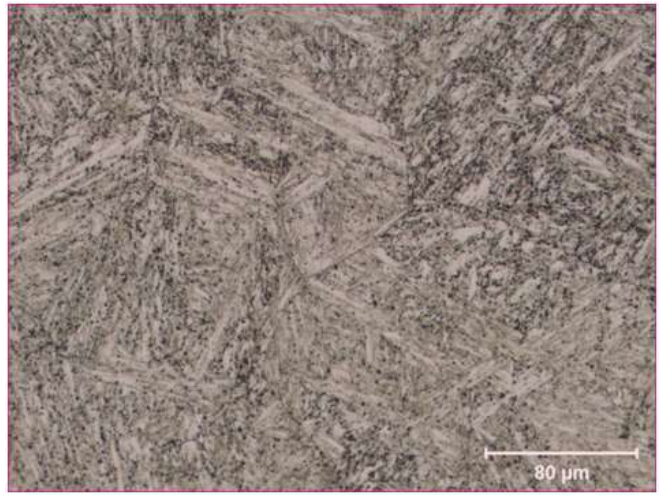

Figure 3 Microstructure following the tempering of the S41500 martensitic stainless steel.

The chemical composition of both base and weld metals is presented in Table **2**. Room temperature tensile properties of S41500 and E309L alloys were characterized according to ASTM E8M standard (ASTM, 2011). They are presented in Table **3**. The S41500 tensile tests were performed in L-T direction, that is, normal to the crack plane using small-size specimen with a 9 mm diameter and a gage length five times the diameter. The head speed was imposed at 5 mm/min. Fracture toughness of S41500 in L-T direction was measured by Chen and his coworkers (Chen, Verreman, Foroozmehr, & Lanteigne, 2013). They found $K_{IC}$ = 316 MPa$\sqrt{\text{m}}$.

|        | **C** | **Si** | **Mn** | **P** | **S** | **Cr** | **Mo** | **Ni** | **N₂** | **Cu** |
|--------|-------|--------|--------|-------|-------|--------|--------|--------|--------|--------|
| S41500 | 0.026 | 0.34 | 0.74 | 0.021 | 0.001 | 13.02 | 0.56 | 3.91 | 0.031 | 0.15 |
| ASTM[2] | <0.05 | <0.6 | 0.5-1 | < 0.03 | <0.03 | 11.5-14 | 0.5-1 | 3.5-5.5 | ... | ... |
| E309L | 0.03 | 0.72 | 2.12 | 0.006 | 0.001 | 24.5 | 0.3 | 13.7 | ... | ... |
| AWS[4] | < 0.03 | 0.3-0.65 | 1.0-2.5 | < 0.03 | < 0.03 | 23-25 | < 0.75 | 12-14 | ... | ... |

Table 2: As-received and nominal composition limits (weight %) of: (i) S41500 alloy used to machine the rigid frame and medallion of the CT-RES specimen; and (ii) 309L alloy used as the filler metal for welding the medallion to the rigid frame

|        | $E$ (GPa) | $\sigma_{ys}$ (MPa) | $\sigma_{ult.}$(MPa) | $A$ (%) |
|--------|-----------|---------------------|----------------------|---------|
| S41500 | 195 | 660 | 822 | 32 |
| E309L | 200 | 358 | 644 | 63 |

Table 3: Room temperature tensile properties of S41500 and 309L alloys

### 2.2 Fabrication of the CT-RES specimen

The fabrication of the CT-RES specimen used the robotic arm Scompi® developed at Hydro-Québec Research Institute (Hydro-Québec, 2011). The actual setup designed to weld the CT-RES specimen is shown in Figure 4. Two torches, one on each side of the specimen, were fixed on a rigid assembly while the robotic arm moved the specimen during the welding process. It is believed that simultaneous deposition of beads on each side of the specimen would significantly reduce distortions and lead to a more symmetric residual stress field in the specimen.

Flux-cored arc welding (FCAW) was used to perform weld deposition between the CT-RES medallion and its rigid frame. The protection gas was constituted of 75% argon and 25% carbon dioxide. A single bead on each side of the medallion, top and bottom, was sufficient to assemble the CT-RES sample. That is a total of four beads. The welding parameters are present in Table 4.

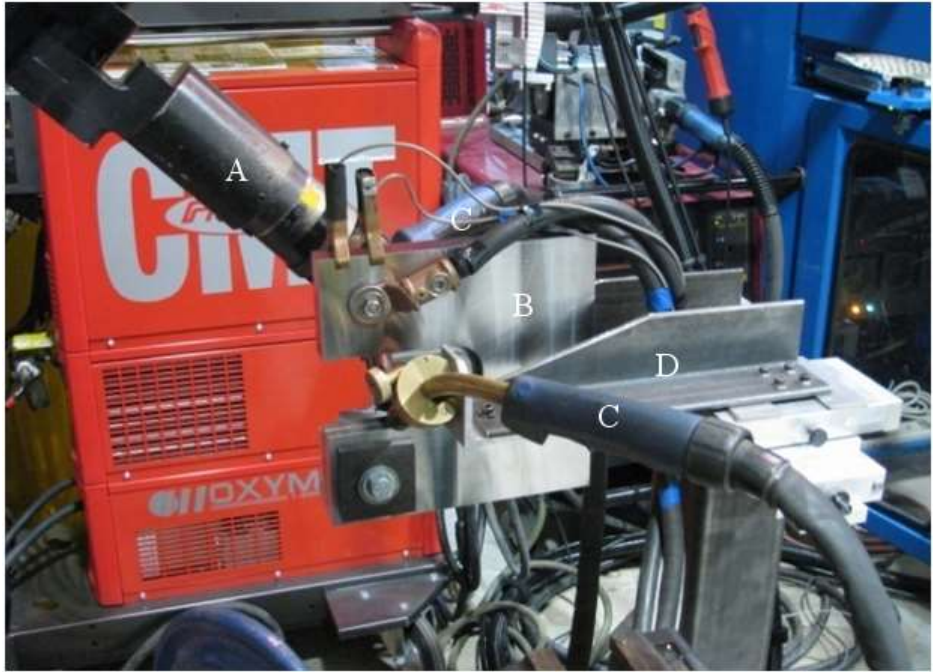

Figure 4: Setup used to weld the CT-RES specimen: (A) Scompi® Robotic Arm; (B) CT-RES specimen; (C) welding torches; and (D) rigid assembly

| Voltage (V) | Current intensity (A) | Wire feed speed (mm/s) | Travel speed (mm/s) | Deposition rate (kg/h) | Linear energy (J/mm) | Preheat temperature (C°) | Interpass temperature (C°) |
|---|---|---|---|---|---|---|---|
| 26 | 170 | 132 | 5.5 | 3.3 | 804 | 100 | 100-115 |

Table 4: Welding parameters used to assemble the CT-RES medallion to the frame

During the process, particular attention was given so that the temperature reached in the medallion did not exceed 630 °C at weld interpass. This was done to avoid the austenite-martensite ($\gamma$-$\alpha'$) phase transformation that occurs in this alloy at 300 °C upon cooling. This phase transformation results in volume increase of approximately 3% that could significantly alter the final residual stress field in the medallion after the assembly. This expansion is clearly visible on a Thermo-Mechanical Analyzer (TMA) curve obtained in a previous study (Lanteigne, Sabourin, Bui-Quoc, & Julien, 2008). Maximum temperature in the medallion never exceeded 350 °C, so that the austenite-martensite transformation was avoided.

## 2.3 Fatigue crack propagation test

The characterization of crack propagation rate was done according to the ASTM E647 standard (ASTM, 2013). Using a 250 $kN$ servo-hydraulic test machine, tests were performed at a fixed frequency of 6 Hz. Prior to the FCP test, an initial stress intensity factor range of $\Delta K$ = 11 MPa$\sqrt{\text{m}}$ was imposed for pre-cracking the samples. The pre-cracking stage was maintained until the crack lengths observed with microscope on each side of the specimen were comparable to within approximately 200 µm.

During the FCP test, the crack length was calculated by the compliance method. For this purpose, a crack mouth opening displacement (CMOD) extensometer was mounted on the medallion. The crack lengths calculated with this method were frequently compared with optical measurements to ensure their validity. Crack length data was automatically sampled at each 250 µm crack advance and the FCPR was then calculated using the secant method defined by the following equation:

$$\frac{da}{dN_a} = \frac{(a_{i+1} - a_i)}{(N_{i+1} - N_i)} \qquad (1.)$$

As for the $\Delta K$, it was calculated using the average crack length of the $da$ interval. Fatigue crack closure was measured using the $P$-$v$ curves. To achieve this, a 4% deviation criterion based on a linear regression fitted to the linear segment (i.e. fully open crack) of the $P$-$v$ diagram was used to determine the closure load, $P_{cl}$, as show in Figure **12**.

To delineate the influence of welding RS on the fatigue crack propagation behavior of the CT-RES specimen, the experimental plan shown in Table **5** is undertaken. The choice made for the $R$ ratios used to perform the FCP test in the CT-Monoblock specimen, i.e. without RS, is based on the actual loading condition present in a turbine runner. In fact, there are two distinct load signatures. As the production begins, the stress measured in the blades rapidly increases to a steady-state value. This load signature is characterized by a low cyclic stress ratio ($R = 0.1$). When the wicket gates are fully open, the hydraulic dynamic load is combined to the static load from which a high cyclic stress ratio ($R = 0.7$) is generated (Sabourin, Bouffard, & Paquet, 2007).

| Specimen | $\Delta K$ [MPa$\sqrt{\text{m}}$] | $R$ | Residual stresses | $K_{max}$ [MPa$\sqrt{\text{m}}$] |
|---|---|---|---|---|
| CT-RES | 15 | $\left( \dfrac{K_{\min} + K_{res}(a)}{K_{\max} + K_{res}(a)} \right)$ | Yes | $K_{max} = \dfrac{\Delta K}{(1-R)} + K_{res}(a)$ |
| CT-Monoblock | 15 | 0.1/0.7 | No | $K_{max} = \dfrac{\Delta K}{(1-R)}$ |

Table 5: Experimental plan to highlight the effect of residual stresses on FCP in the CT-RES specimen

## 3. Computation of stress intensity factor and compliance

The following equation of linear elastic fracture mechanics (LEFM) applicable to standard CT-specimen geometry is used to correlate the load P to the stress intensity factor :

$$K = \frac{P}{b_m \sqrt{W}} Y(a/W) \qquad (2.)$$

in which the dimensionless solution of $K$, $Y(a/W)$, is a polynomial expansion.

When this equation is applied to the CT-RES specimen, $P$ represents the test force, $b_m$ the thickness of the medallion, $W$ the distance between the loading line and the end of the medallion (*cf.* Fig. 2a.). The geometrical factor $Y(a/W)$ had to be calibrated numerically by the finite element method (FEM). To ensure an adequate calibration of the CT-RES geometry, three methods are used to calculate its SIF: (i) the $J$-Integral method (Rice, 1967); (ii) the displacement extrapolation at the crack tip; and (iii) the equation of an edge crack in a semi-infinite sheet. It can be noted that methods (i) and (ii) are available in the finite element (FE) software Ansys®. The FE model used represents a quarter of the entire geometry. Symmetry boundary conditions were applied on the crack propagation plane and on the mid-thickness plane.

The implementation of the first method ($J$-Integral) requires multiple calculations. In each of these calculations, a crack of length $a$ is introduced by changing the boundary conditions on the crack plane. The mesh surrounding the crack tip is constituted of 20-nodes quadratic hexahedral elements. The convergence of the method is verified over 10 paths to insure a good validation.

In the second technique, the mesh is modified to respect the square root distribution of displacements near the crack tip. Thus, the second node of each element defining the crack front is moved at $L/4$ from the crack tip, as proposed by Barsoum (Barsoum, 1977). The parameter $L$ is the element length.

The third method uses the analytical equation giving the stress intensity factor of an edge crack in a semi-infinite sheet to calculate $K$ for mechanically short cracks ($a/W < 0.5$) in the CT-RES specimen. Hence,

$$K = 1.1215\ \sigma_{nom}\sqrt{\pi a^*} \qquad (3.)$$

where $\sigma_{nom}$ represents the stress component $\sigma_y$ on the medallion side of the un-cracked CT-RES geometry subjected to the same load $P$ used to calibrate the cracked geometry.

Results obtained with the three methods are presented in Figure **5**.

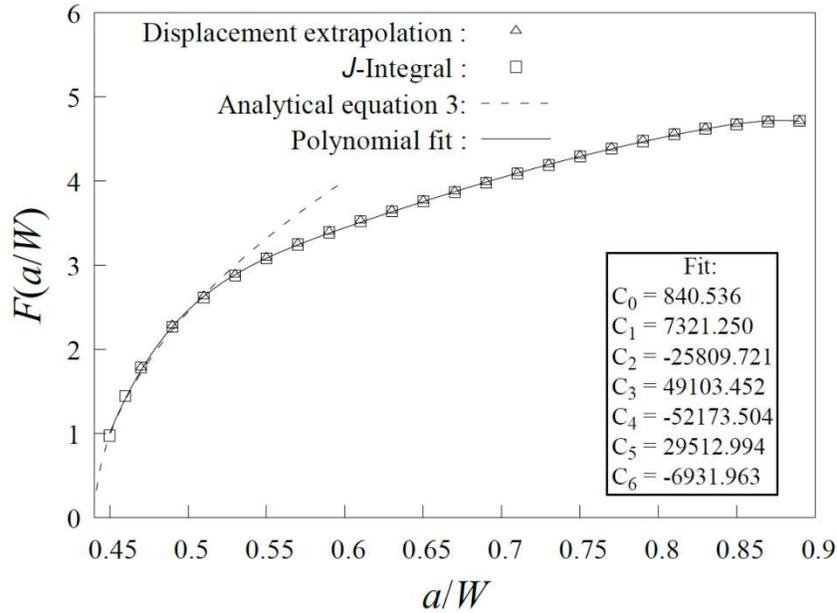

Figure 5: Dimensionless solution of $K$ computed for the CT-RES and CT-Monoblock specimens with: the $J$-Integral method; the displacement extrapolation technique; and the analytical equation of a crack edge in a semi-infinite sheet.

Figure **5** show a good agreement between the three different methods. The analytical equation, corresponding to the dashed line of the figure, is valid for crack lengths up to $a/W = 0.52$. For convenience, the function $Y\ (a/W)$ calculated in this study was obtained from a polynomial fit of the $J$-Integral results. The polynomial fit equation is defined as:

$$Y(a/W) = \sum_{i=0}^{6} C_i \cdot (a/W)^i \qquad (4.)$$

The CT-RES specimen compliance has also been calibrated by the FEM. Crack mouth opening displacements were calculated for a series of crack lengths obtained by incrementing successively $a/W$ by a value of 0.02. The dimensionless compliance values, $v_x$, were then calculated using the equation proposed by Saxena (Saxena & Hudak, 1978):

$$v_x = \left\{ \left( \sqrt{\frac{Evb_m}{P}} \right) + 1 \right\}^{-1} \qquad (5.)$$

where $v$ corresponds to the crack mouth opening displacement and $E$ the elastic modulus. The data obtained from this method is then smoothed using a polynomial regression, the results of which are used to calculate the crack length during the FCP test. The polynomial function best representing the compliance behavior of the CT-RES specimen is shown in Figure **6** and is defined as:

$$(a/W) = \sum_{i=0}^{6} C_i \cdot v_x{}^i \qquad (6.)$$

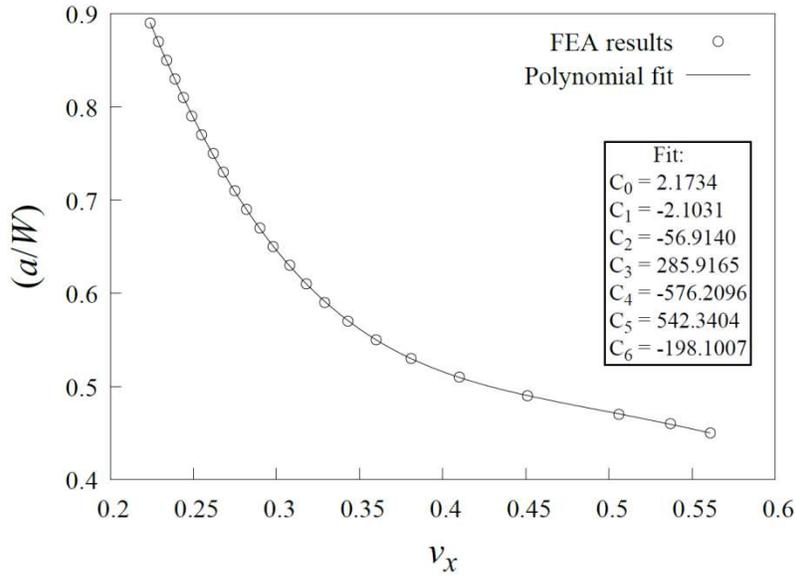

Figure 6: Dimensionless compliance values computed from FE analysis for the CT-RES and CT-Monoblock specimens

It is noteworthy that the difference between visual measurements and crack lengths calculated with the compliance method never exceeds 0.5 mm. Since the initial crack length is 37.5 mm, it follows that the maximum error introduced by the compliance method is approximately 1.3% on the crack length. The precision of the compliance method is thus sufficient to avoid significant errors on the applied loads.

## 4. Residual stresses in the CT-RES specimen

### 4.1 Measurement of the residual stress field

The residual stress field introduced in the CT-RES specimen after welding was measured by the contour method developed by Prime (Prime, 2001). This technique allows the determination of the distribution of the normal component of an unknown residual stress field by cutting the specimen along a plane perpendicular to that stress component (Levesque, 2015).

This method is used herein to calculate the distribution of the residual stress component $\sigma_y^{res}$ that is perpendicular to the crack propagation plane. To begin, the CT-RES specimen is cut along the crack propagation plane by electrical discharge machining (EDM). Then, an optical profilometer is used to measure the relaxed normal displacements along the two freshly created surfaces. Finally, the average displacement map is imposed as boundary conditions on a FE model of the CT-RES specimen to compute the corresponding residual stress distribution.

However, modification of the original displacement maps obtained with the profilometer is required before calculating the residual stress field. First, there is an abrupt increase in displacement at the very end of the cut, seemingly caused by plastic deformation. The displacements in this plastically deformed region were replaced by a 3D linear extrapolation from the rest of the displacement distribution. Second, the results obtained near the edges (≈1 mm) were discarded because it is well known that they are inaccurate (Hosseinzadeh, Ledgard, & Bouchard, 2012; Sarafan, Lévesque, Wanjara , Gholipour, & Champliaud, 2015). It is also important to note that the contour method is a destructive procedure. As a matter of fact, two CT-RES specimens were fabricated. The first one was used for the measurement of the residual stress field in the medallion, while the second was used for the FCP test. Therefore, the analysis relies on the hypothesis that both samples have identical residual stress fields. Figure 7 illustrates a distribution of the residual stress component $\sigma_y^{res}$ at mid-thickness of the medallion and calculated by the contour method. The oscillations are caused by the smoothed displacement map obtained from cubic splines. The variation of stress across the thickness is not significant.

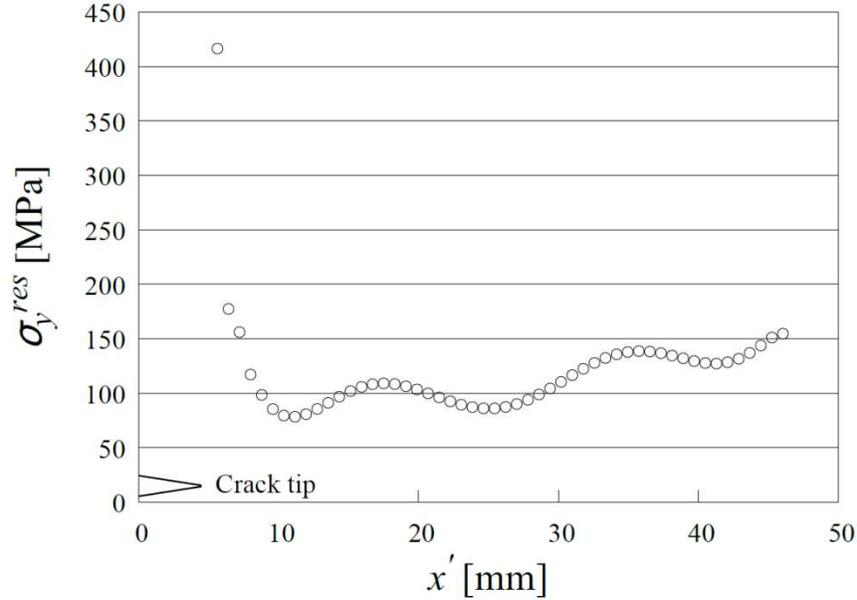

Figure 7: Mid-thickness distribution of $\sigma_y^{res}$ normal to the crack propagation plane in the medallion of the CT-RES specimen.

It can be seen in Figure 7 that no residual stresses are captured for the first 5 mm due to the pre-cracking of the specimen. This pre-crack also led to a significant increase of the residual stress ahead of the crack tip as observed by Liljedahl et al.,(Liljedahl et al., 2010)and Trudel et al. (Trudel, Sabourin, et al., 2014).

The results in Figure 7 also show a continuous increase in the tensile residual stresses along the medallion width (x'). This is different from the residual stress profiles in self-equilibrated specimens. This particular feature of the CT-RES makes it a unique specimen for studying the influence of RS on FCPR.

To get an idea of the behavior of the residual stress field redistribution as the crack propagates in the medallion, a finite element modeling was performed to recreate thermal contraction induced by the welding of the medallion in the rigid frame. In the simulation, elements within the weld deposit were contracted artificially. The induced stress in the uncracked planed, $\sigma_f$, was computed as the sum of all reaction forces, $f_i$, normal to the crack plane divided by the area of the remaining ligament, $A$. This simple equation is defined as followed:

$$\sigma_f = \frac{\Sigma f_i}{A} \qquad (7.)$$

Figure **8** presents the evolution of the normalized mean stress in the remaining ligament as the crack propagates. We can see that the mean stress constantly increases as the remaining ligament of the medallion becomes shorter. That short analysis informs us on the behavior of the $\sigma_y^{res}$ redistribution and thus, the evolution of $K_{res}$ in the CT-RES specimen. It is important to specify that this simulation does not recreate the real conditions of the welding procedure but therefore give a tendency of the redistribution of the RS in the CT-RES medallion as the crack propagates.

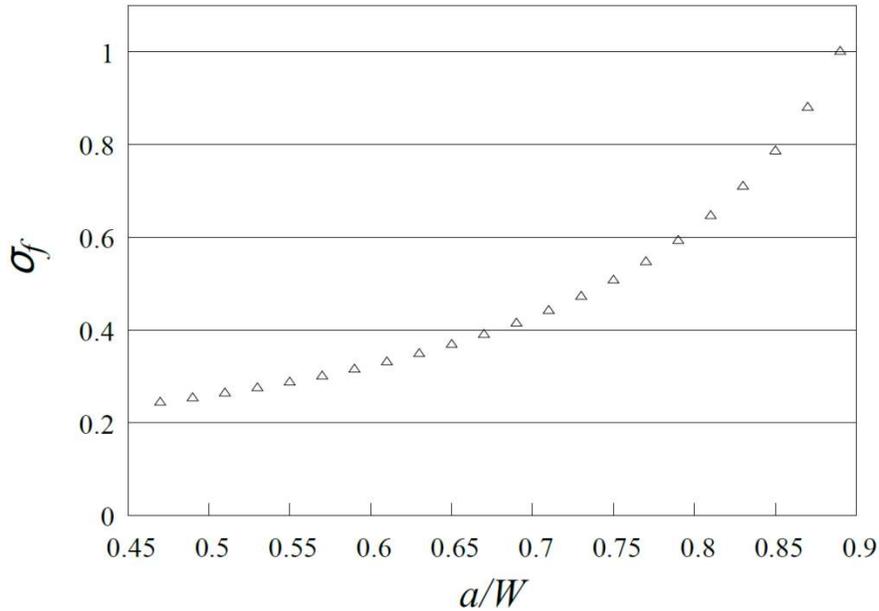

Figure 8: Evolution of the mean stress $\sigma_f$ in the remaining ligament of the medallion as the crack propagates

Using this novel specimen, our experimental strategy is to perform crack propagation tests at constant $\Delta K$ and $R$ ratio to isolate the effect of residual stresses on crack growth behavior. By maintaining constant $\Delta K$ during a fatigue crack propagation test, $\sigma_{res}$ remains the only varying parameter, hence the influence of RS on fatigue crack propagation can be studied while avoiding the effect of any other parameter variation.

### 4.2 Computation of $K_{res}$ with the weight functions method

The variation of $K_{res}$ with the advance of the crack in the CT-RES specimen was analyzed using the weight functions method developed by Bueckner (Bueckner, 1970). This method allows the calculation of the SIF resulting from any residual stresses distribution and can be used with a wide variety of specimen geometries.

### 4.2.1 Weight function theory

Rice (Rice, 1972) proposed a simplified expression to calculate the stress intensity factor, $K_{I,II}$, in a 2D cracked structure submitted to surface traction $\boldsymbol{t}$ on a boundary $\Gamma$ and submitted to body forces $\boldsymbol{f}$:

$$K_{I,II} = \oint \boldsymbol{t} \cdot \boldsymbol{h} \, d\Gamma + \int \boldsymbol{f} \cdot \boldsymbol{h} \, dV \qquad (8.)$$

where $\boldsymbol{h}$, the weight function, is defined as:

$$h(x, y, a) = \frac{E}{K} \frac{\partial \boldsymbol{u}(x, y, a)}{\partial a} \qquad (9.)$$

The second term of the Eq.8 is discarded in this work because the CT-RES specimen is not submitted to body forces. Moreover, the two-dimensional function $\boldsymbol{h}$ is reduced to a one dimension function, $h(x, a)$, since the calibration of the method was done only on the $u_y$ component ($K_I$) of the global displacement field, $\boldsymbol{u}$. The final form of the equation can be defined as:

$$K_I = \int_{a_0}^{a} \sigma_y^{res}(x) \; h(x, a) \; dx \qquad (9.)$$

The work of Glinka & Shen (Glinka & Shen, 1991)was exploited in order to fit the displacement derivative $\partial u_y / \partial a$ with the universal equation developed by the authors, Eq.10, composed of $M_i$ unknown coefficients. For a deep edge crack, which is assumed in the case of the CT-RES specimen, the work of Fett et *al.* (Fett, Mattheck, & Munz, 1987)also contributes to the simplification of the calibration since $M_2 = 3$. Hence, the universal function is written as:

$$h(x, a) = \frac{2}{\sqrt{2\pi(a-x)}}\left[1 + M_1\left(1 - \frac{x}{a}\right)^{\frac{1}{2}} + 3\left(1 - \frac{x}{a}\right) + M_3\left(1 - \frac{x}{a}\right)^{\frac{3}{2}}\right] \qquad (10.)$$

### 4.2.2 Weight function calibration

To perform the calculation of $M_i$, two linearly independent stress distributions must be applied on the CT-RES specimen boundary as proposed by Shen & Glinka (Shen & Glinka, 1991). They are chosen as: (i) a constant stress distribution; and (ii) a linearly increasing stress distribution from the load line to the end of the frame. These two reference stress distributions were applied to the CT-RES rigid frame in the FE model.

Following the calculation, a validation was executed to ensure the accuracy on the calculation of $K_{res}$. To do so, a volumetric contraction was imposed on the weld bead elements (*cf*. Fig.2b) of the FE model. The *J*-Integral method was then used to determine $K$ for various crack lengths and the results were compared to those obtained by the weight functions method. Both methods showed a very good agreement. Indeed, the largest difference between the SIF calculated by the weight functions method and the one obtained with the *J*-Integral method was less than 0.1%.

### 4.2.2 $K_{res}$ calculation performed by the weight function method

Once validated using known $K_{res}$ distributions, the weight functions method can be used to evaluate the stress intensity factor induced by the CT-RES residual stress field for different crack lengths. The results are shown in Figure **9**. The first point in Figure **9** (the black dot) had to be corrected since the value was too low($K_{res} = 8 \text{MPa}\sqrt{m}$)compared to neighboring values. The value was extrapolated from the two following values. This is probably caused by a lack a precision from the calculation of the stress near the crack tip in the FE model.

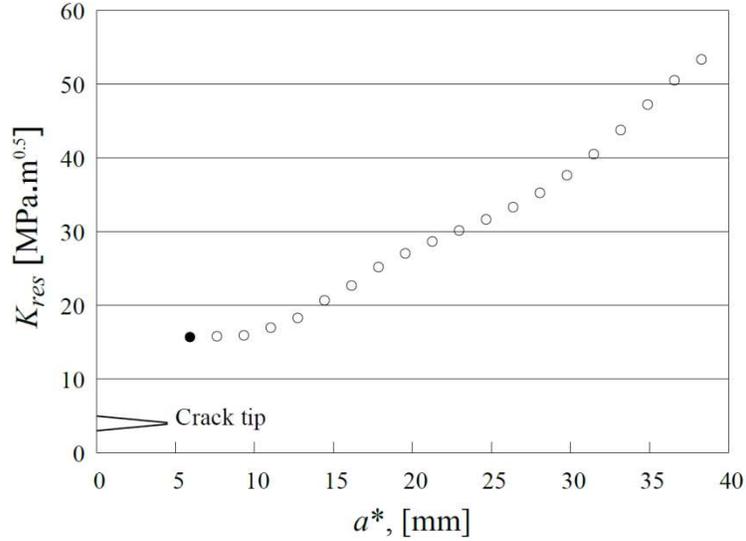

Figure 9: Computed $K_{res}$ from the weight functions method as function of crack advance in the medallion

Figure **8** and Figure **9** respectively show an increase of the residual stress and the residual stress intensity factor $K_{res}$ as the crack progresses into the residual stress field, which is an expected behavior.

## 5. Experimental results

### 5.1 Testing the CT-RES: FCP in a residual stress field

Figure **10** shows the evolution of the minimum and maximum stress intensity factors calculated using the weight functions method plotted against the crack length. The values $K_{min-app}$ and $K_{max-app}$ represent the minimum and maximum applied SIF by the servo-hydraulic bench, known as the external load. The gray band corresponds to the external solicitation i.e $\Delta K$. As mentioned before, the $\Delta K$ was kept constant during the test so as to highlight the potential effect of the RS field on the FCPR in the CT-RES sample. Consequently, the variation of $K_{res}$ along the medallion modifies the minimum and maximum SIFs, i.e. $K_{min}$ and $K_{max}$, while not affecting the difference between the two. This results in a constant $\Delta K$ fatigue crack propagation test in which the local cyclic stress ratio ($R_{local}$) evolves with the crack length from $R = 0.55$ to $R = 0.8$ as shown in Figure **10**.

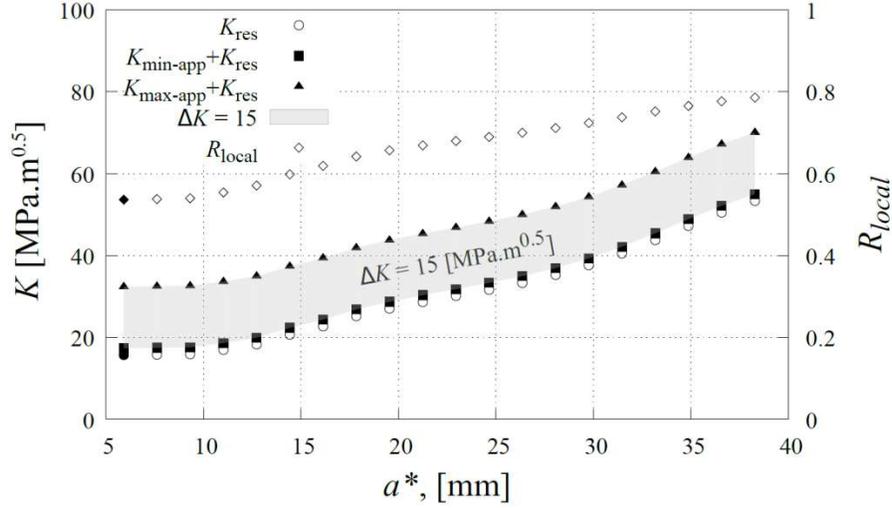

Figure 10: Plot of the variation of SIF in the CT-RES specimen during an FCP test at constant applied $\Delta K$ = 15 MPa$\sqrt{m}$

Figure 11 shows the comparison between fatigue crack propagation rates in the CT-RES and CT-Monoblock specimens at $\Delta K$ = 15 MPa$\sqrt{m}$. The dashed lines correspond to punctual values of FCPR extracted from d$a$/d$N$ - $\Delta K$ curves ($R$=0.1 and $R$=0.7) performed at increasing $\Delta K$ in the Paris regime.

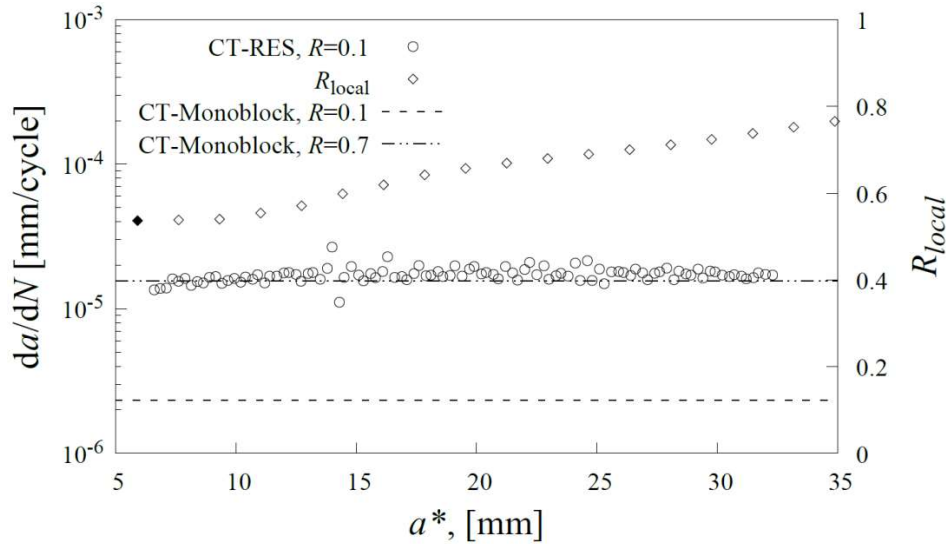

Figure 11: Crack propagation rates in the CT-RES and CT-Monoblock specimens when solicited at $\Delta K$ = 15 MPa$\sqrt{m}$

As seen in Figure **11**, even with an increase of the cyclic stress ratio at the crack tip, $R_{local}$, induced by the residual stress field, the FCPR remains constant during the test. The average rate of crack propagation in both the CT-Monoblock and CT-RES specimen are listed in Table **6**.

| Specimen | Applied $R$ ratio | d$a$/d$N$[mm/cycle] |
|---|---|---|
| CT-Monoblock | 0.1 | 2.33E-06 |
| CT-Monoblock | 0.7 | 1.56E-05 |
| CT-RES | 0.1 | 1.72E-05 |

Table 6: Fatigue crack propagation rate in CT-RES and CT-Monoblock specimens at $\Delta K$ = 15 MPa$\sqrt{m}$

During the FCP test performed on the welded CT-RES specimen, no crack closure was detected as presented by the $P$-$v$ results A and B plotted on Figure **12**a. Series A and B were respectively acquired during the last (32 mm) and the first (6 mm) crack propagation cycles. It is apparent that the tensile RS in the specimen opened the crack, eliminating crack closure. The closure free behavior was confirmed by the perfect superposition of a straight line on the two sets of results. On the other hand, the $P$-$v$ curve C, corresponding to the data recorded during the test performed at $R = 0.1$ in the CT-Monoblock specimen, presents crack closure. The arrow on Figure **12**b indicates the crack closure load, $P_{cl}$.

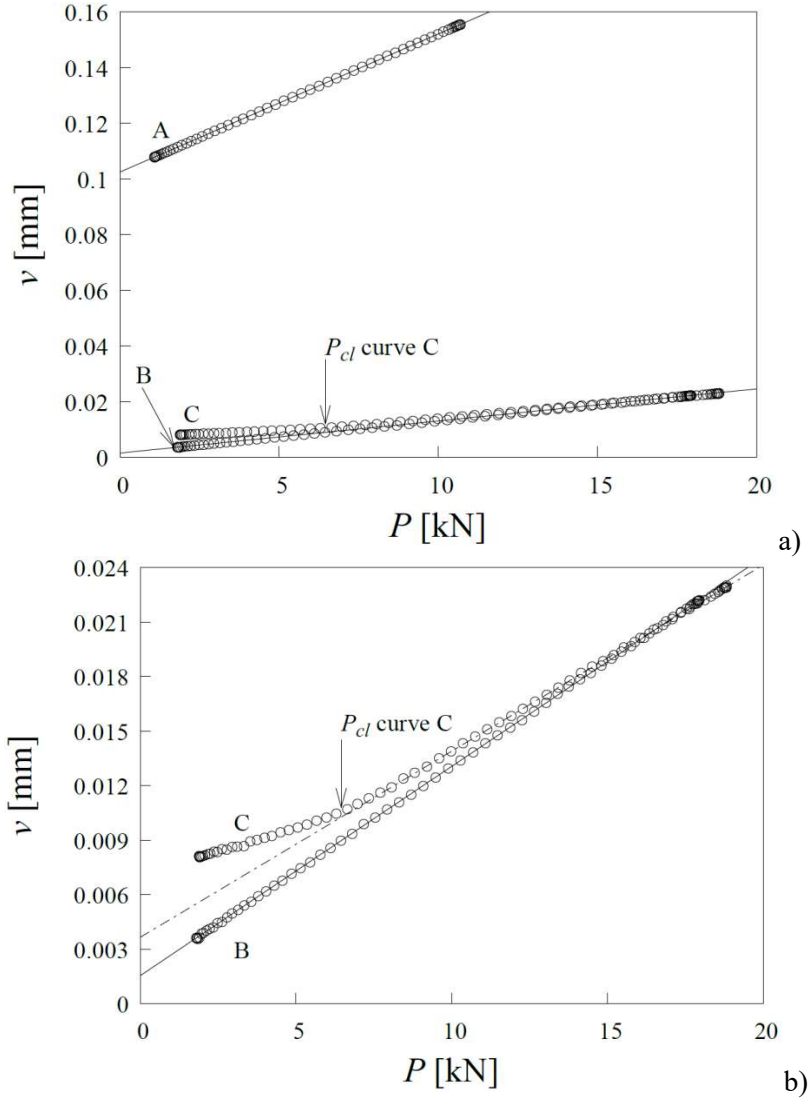

Figure 12: (a) The first (A) and last (B) $P$-$v$ curves monitored in the welded CT-RES specimen, as well as the $P$-$v$ curve (C) monitored in the CT-Monoblock specimen at $\Delta K = 15$ MPa$\sqrt{\text{m}}$ and $R = 0.1$, (b) the curves B and C zoomed to highlight the crack closure load $P_{cl}$

In opposition, the $P$-$v$ results acquired for the CT-Monoblock specimen tested at $R = 0.7$ showed no evidence of crack closure. It is therefore expected that the FCPR behavior of the CT-RES specimen will be closer to the FCPR of the CT-Monoblock tested at $R$=0.7, as shown by the results in Table 6. Nevertheless, a small difference persists between these two behaviors. The average fatigue crack growth rate in the CT-RES is approximately 10% higher than that of the CT-Monoblock specimen tested at $R = 0.7$. Even if this difference could be attributed to result dispersion, (Virkler, Hillberry, & Goel, 1979), another hypothesis points towards the effect of RS on $K_{max}$. For a given $\Delta K$, $K_{max}$ is known to influence the fatigue crack growth rate in certain materials. For example, aluminum alloys are more sensitive to an increase in $K_{max}$, than mild steels or stainless steels (Koçak, Webster, Janosch, Ainsworth, & Koers, 2006).In fact, the low values of $K_{IC}$ in aluminum alloys is penalizing since a low value of $K_{max}$ could triggered static damage mechanics and increase the FCPR. It is thus proposed to briefly analyze our material sensitivity to $K_{max}$, using the Nasgro model.

In the NASGRO model(Forman, R. G. & Mettu, 1992), the sensitivity of an alloy to $K_{max}$ is quantified by an exponent $q$. The NASGRO equation giving the crack growth rate as a function of the applied SIF range $\Delta K$, maximum SIF $K_{max}$, cyclic stress ratio $R$ and the effective stress ratio $f$, function of $R$, as defined by Newman 6 is:

$$\frac{da}{dN} = C\left[\left(\frac{1-f}{1-R}\right)\Delta K\right]^n \frac{\left(1 - \frac{\Delta K_{th}}{\Delta K}\right)^p}{\left(1 - \frac{K_{max}}{K_{IC}}\right)^q} \tag{11.}$$

In this equation, parameters $C$, $n$, $p$, and $q$ are material constants that must be determined experimentally. For FCP tests performed with a positive $R$, the parameter $q$ is equal to 1 for the majority of aluminum alloys and 0.25 for the majority of stainless steels 7(Forman, R.G. , Shivakumar, Cardinal, William, & McKeighan, 2005; Koçak et al., 2006).

The NASGRO model is used herein to estimate the influence of $K_{max}$ on the FCPR of the CT-RES specimen. Since the ratio containing both $R$ and $f$ is equal to 1 during the FCP test (crack fully open), the analysis is limited to the term quantifying the sensitivity of this alloy to $K_{max}$. The coefficients $p$ and $q$ were assumed to be equal to 0.25 due to the lack of data for SS415 stainless steel in the literature.

To highlight the influence of $K_{max}$ on the FCPR, the starting and ending values of $K_{max}/K_{IC}$ ratio obtained during FCP in the CT-RES specimen are presented in Table 7.

|  | Starting | Ending | Difference |
|---|---|---|---|
| $K_{max}/K_{IC}$ | 0.101 | .190 | + 46 % |
| $\left(1 - \frac{K_{max}}{K_{IC}}\right)^{0.25}$ | 0.973 | 0.948 | - 2.57 % |

Table 7: Evaluation of the effect of $K_{max}/K_{IC}$ ratio on the predicted FCPR in CT-RES specimen by the NASGRO model

As we can see from Table 7, a significant increase of the ratio $K_{max}/K_{IC}$ generates a slight decrease of the denominator in the NASGRO model, Eq.11. Such a small difference is negligible relative to the results dispersion.

This entails that stainless steels suffer no particular effect from residual stresses, other than a *R effect*, with respect to FCPR for the tested loading conditions. Consequently, a tensile residual stress field mainly influences the FCPR by preventing crack closure as would high stress ratio do. These observations partially validate the philosophy of the BS7910 standard 8 (BSI, 2013)recommending that the remaining life of a cracked structured subjected to tensile RS should be calculated from propagation curves characterizing fully opened cracks ($R \geq 0.5$). However, this philosophy could lead to significant underestimation of the remaining life in the case of partial or complete closure of the crack. Thus, the design optimization and improvement in fatigue life prediction of structures facing dynamic load will inevitably require broad knowledge of both metallurgical and mechanical notions. To name a few, a thorough comprehension of the effect of phase transformation in materials, the calculation of the residual stress field in assembled structures by multiphysics FE simulation and the effect of multiaxial, proportional or non-proportional, loading on the damage nucleation and crack propagation.

## 6. Conclusion

In this study, a novel specimen geometry was developed to study the influence of welding residual stresses RS on fatigue crack propagation in SS415, while avoiding the influence of other factors such as microstructural changes or plastic deformations. The CT-RES specimen was manufactured by welding a medallion in a rigid frame. Both the medallion and rigid frame were composed of SS415 stainless steel. The resulting residual stress field is believed to be very similar to the RS field observed in turbine runners after fabrication. The crack propagation rate measured in the welded specimen is nearly identical to the crack propagation rate observed in the CT-Monoblock specimen, i.e. without RS, at an applied $R$ ratio of 0.7. From the findings presented in this study, the following conclusions may be drawn:

- The CT-RES specimen can be used to study the influence of welding residual stress on FCPR;
- The CT-RES can be used for testing different welding procedures as well as the effect of the following post-weld treatment on the RS level;
- The fabrication of the CT-RES specimen does not alter the material microstructure in the FCP region of the specimen, or its consolidation state. In other words, this sample can be used to isolate the effect of welding RS on fatigue crack propagation behavior;
- The absence of crack closure accounts for most of the crack propagation rate increase observed in the welded CT-RES specimen. This phenomenon is directly related to the presence of tensile residual stresses resulting from the welding process;
- The redistribution of the residual stress field in the constrained medallion has no effect on propagation as long as the tensile residual stress magnitude is high enough to fully open the crack lips. As a consequence, crack propagation rates measured in the CT-RES and CT-Monoblock specimens solicited at $R = 0.7$ are very similar.

### Acknowledgments

The authors gratefully acknowledge CReFaRRE, General Electric, Hydro-Québec, CRSNG, MITACS for their financial support. P.-A. Deschênes would also like to acknowledge Carlo Baillargeon, Manon Provencher, René Dubois, Yvan Laroche and Jean-Benoît Hudon for their technical help during the project.